\documentclass[prl,twocolumn,fleqn]{revtex4}
\pdfoutput=1
\usepackage{epsfig}
\usepackage{epstopdf}
\usepackage{graphicx}

\usepackage{amsmath}
\usepackage[psamsfonts]{amssymb}
\usepackage{euscript}

\usepackage{latexsym}

\def\hsp{,\hspace{.7cm}}

\def\thet#1#2{\theta_{#1#2}}

\def\spp#1#2#3#4{\hbox{\rm sin}^#1(#2\thet#3#4)}

\renewcommand{\sin}{\textrm{sin}}

\renewcommand{\(}{\begin{equation}}
\renewcommand{\)}{end{equation} \vspace{-.05in}\linebreak}

\newcounter{saveeqn}
\newcounter{savealpheqn}

\newcommand{\alpheqn}{\setcounter{saveeqn}{\value{equation}}%
  \stepcounter{saveeqn}\setcounter{equation}{0}%
  \renewcommand{\theequation}{\mbox{\arabic{section}.\arabic{saveeqn}
\alph{equation}}}
  \renewcommand{\)}{\end{equation}}}
\def\part#1{\frac{\partial}{\partial{#1}}}%
\def\group#1{\refstepcounter{equation}\setcounter{saveeqn}
 {\value{equation}}%
  \label{#1}\setcounter{equation}{0}%
\renewcommand{\theequation}{\mbox{\arabic{section}.\arabic{saveeqn}
\alph{equation}}}
  \renewcommand{\)}{\end{equation}}}
\newcommand{\reseteqn}{\setcounter{equation}{\value{saveeqn}}%
  \renewcommand{\theequation}{\arabic{section}.\arabic{equation}}%
  \renewcommand{\)}{\end{equation}}}

\newcommand{\aalpheqn}{\setcounter{saveeqn}{\value{equation}}%
  \stepcounter{saveeqn}\setcounter{equation}{0}%
  \renewcommand{\theequation}{\mbox{
        \Alph{subsection}.\arabic{saveeqn}\alph{equation}}}
   \renewcommand{\)}{\end{equation}}}
\newcommand{\areseteqn}{\setcounter{equation}{\value{saveeqn}}%
  \renewcommand{\theequation}{\Alph{subsection}.\arabic{equation}}%
  \renewcommand{\)}{\end{equation}}}

\renewcommand{\thefootnote}{\alph{footnote}}
\renewcommand{\(}{\begin{equation}}
\renewcommand{\)}{\end{equation}}
\newcommand{\ba}{\begin{eqnarray}}
\newcommand{\ea}{\end{eqnarray}}

\newcommand{\bp}{\mathop{\vtop{\ialign{##\crcr
   $\hfil\displaystyle{}\hfil$\crcr\noalign{\kern-13pt\nointerlineskip}
   \BIG{(}\hskip0pt\crcr\noalign{\kern3pt}}}}}
\newcommand{\cbp}{\mathop{\vtop{\ialign{##\crcr
   $\hfil\displaystyle{}\hfil$\crcr\noalign{\kern-13pt\nointerlineskip}
   \BIG{)}\hskip0pt\crcr\noalign{\kern3pt}}}}}
\newcommand{\pa}{\mathop{\vtop{\ialign{##\crcr
    
$\hfil\displaystyle{\oplus}\hfil$\crcr\noalign{\kern+1pt\nointerlineskip 
}
   \hspace{.08in}$^{\alpha=0}$\hskip6pt\crcr\noalign{\kern3pt}}}}}
\renewcommand{\hsp}{,\hspace{.3in}}

\newcommand{\beq}{\begin{equation}}
\newcommand{\eeq}{\end{equation}}

\numberwithin{equation}{section}
\renewcommand{\theequation}{\mbox{\arabic{equation}}}

\def\hsp#1{\hspace{#1in}}

\catcode`\@=11
\def\vereq#1#2{\lower3pt\vbox{\baselineskip1.5pt \lineskip1.5pt
\ialign{$\m@th#1\hfill##\hfil$\crcr#2\crcr\sim\crcr}}}
\catcode`\@=12

\makeatletter
\newcommand\figcaption{\def\@captype{figure}\caption}
\newcommand\tabcaption{\def\@captype{table}\caption}
\makeatother
\renewcommand{\(}{\begin{equation}}
\renewcommand{\)}{\end{equation}}

\renewcommand{\beq}{\begin{equation}}
\renewcommand{\eeq}{\end{equation}}
\newcommand{\bea}{\begin{eqnarray}}
\newcommand{\eea}{\end{eqnarray}}
\newcommand{\beas}{\begin{eqnarray*}}
\newcommand{\eeas}{\end{eqnarray*}}

\newcommand{\bquo}{\begin{quote}}
\newcommand{\enqu}{\end{quote}}

\def\e{\epsilon}

\def\hsp{,\hspace{.2cm}}

\begin{document}
\def\thefootnote{\fnsymbol{footnote}}

\title{Neutrino Physics with Accelerator Driven Subcritical Reactors}

\author{Emilio Ciuffoli, Jarah Evslin and Fengyi Zhao}
\affiliation{ Institute of Modern Physics, CAS, NanChangLu 509, Lanzhou 730000, China
}

\begin{abstract}
\noindent

\noindent
Accelerator driven system (ADS) subcritical nuclear reactors are under development around the world.  They will be intense sources of free, 30-55 MeV $\mu^+$ decay at rest $\overline{\nu}_\mu$.  These ADS reactor neutrinos can provide a robust test of the LSND anomaly and a precise measurement of the leptonic CP-violating phase $\delta$, including $sign({\rm{cos}}(\delta))$.   The first phase of many ADS programs includes the construction of a low energy, high intensity proton  or deuteron accelerator, which can yield competitive bounds on sterile neutrinos.  




\end{abstract}

%
\setcounter{footnote}{0}
\renewcommand{\thefootnote}{\arabic{footnote}}


\maketitle

With their ability to enrich spent fuel and their inability to melt down, accelerator driven system (ADS) subcritical reactors are being designed and prototypes are being built around the world.   ADS machines in general will be powered by proton beams, of energies of 1-2 GeV with currents of 10-20 mA, which strike high $Z$ targets.   Beam energies of 400 MeV or more lead to the production of pions.  The $\pi^-$ are generally absorbed in the target, while $\pi^+$ stop and decay at rest (DAR), producing $\mu^+$ which also stop and decay at rest, producing $\overline{\nu}_\mu$ with a known energy spectrum peaked between 30 and 55 MeV.  Like the $\overline{\nu}_e$ from ordinary critical reactors, these $\overline{\nu}_\mu$ are free byproducts of the reaction, requiring no modification to the reactor design and so are at no cost to the neutrino physicist.  Early phases of ADS use less energetic beams which can produce $\overline{\nu}_e$ via the decay at rest of spallation isotopes (IsoDAR).   In this letter we describe some of the neutrino physics which can be done with these ADS reactor neutrinos.


For example, in January 2015 the Belgian government approved the MYRRHA installation at a price of 1.5 billion euros, whose construction will begin already in 2017 and finish in 2021.  It is an ADS research reactor with a maximum proton energy of 600 MeV and current of 4 mA, and so in particular it will already be a powerful source of $\mu$DAR $\overline{\nu}_\mu$.

For concreteness we will consider two configurations which are currently planned by the Chinese ADS (C-ADS) collaboration.  The first is a prototype called CI-ADS which consists of a deuterium beam with 100 MeV/nucleon and a current of 15 mA.  The second, corresponding to C-ADS phase II, is an 800 MeV proton beam with a current of 10 mA.  As the beam energy in the first setup is below 400 MeV, it cannot provide $\mu$DAR $\overline{\nu}_\mu$ and so will be used for a $\overline{\nu}_e$ program which will constrain the sterile neutrino mixing angle $\theta_{ee}$.  The second will use $\overline{\nu}_\mu$ to constrain the sterile neutrino mixing angle $\theta_{\mu e}$, providing a robust test of the LSND anomaly, and to measure the leptonic CP-violating phase $\delta$.  

These experimental goals and their implementations are inspired by the DAE$\delta$ALUS project~\cite{daed}.  However there are several key differences.  DAE$\delta$ALUS uses multiple accelerators with various baselines to the detector.  To determine which event came from which source, it requires that only one source runs at a time.  Therefore the maximum instantaneous current at each DAE$\delta$ALUS source needs to be quite high, in the case of the far source they envisage 30-50 mA.  On the other hand, ADS projects consist of a single source and so the beam can be truly continuous and the same number of events/baseline can be achieved with a much lower maximum instantaneous current.  Second, DAE$\delta$ALUS proposes the construction of its own cyclotrons, in fact this is necessary since no country's ADS program envisage the construction of multiple machines with the 10-20 km separation required by their proposal.  Thus in a DAE$\delta$ALUS program, neutrino physicists need to pay for the neutrinos.  Third, as a commercial reactor ADS has a highly redundant design.  Even the C-ADS injecting accelerators are redundant.  This redundancy will allow ADS to have very little downtime.  Finally, in the case of C-ADS, the location in southern China has the advantage that the horizontal component of the Earth's magnetic field is 0.38 G, as compared with 0.17 G at the original DAE$\delta$ALUS location in South Dakota and 0.13 G at the second DAE$\delta$ALUS location in the Pyh\"{a}salmi mine.  As a result the main backgrounds, which arise from low energy atmospheric neutrinos, are reduced by about a factor of two in our setup.

In the first experimental setup, which is similar to that proposed in Refs.~\cite{isodarussi2005,isodar}, we use the 100 MeV/nucleon deuterium beam to knock neutrons out of ${}^9$Be.  These neutrons are moderated and multiplied by heavy water moderator and then absorbed by a 99.99\% isotopically pure ${}^7$Li, producing ${}^8$Li.  The eventual $\beta$ decay of the ${}^8$Li produces our $\overline{\nu}_e$ with an energy spectrum peaked at 6 MeV but extending up to 13 MeV.  These $\overline{\nu}_e$ travel isotropically and are eventually detected via the inverse beta decay reaction
\beq
\overline{\nu}_e + p^+ \rightarrow n + e^+
\eeq
on Hydrogen in liquid scintillator detectors placed at distances between 5 and 30 meters.  Neutrino oscillations between the three known flavors are inappreciable at distances below 100 meters, thus a deficit of detected $\overline{\nu}_e$ suggests oscillations to sterile neutrinos.  

Backgrounds are notoriously problematic in such experiments, and so it is essential to place multiple detectors at multiple baselines.  The baseline-dependence of the neutrino disappearance is well-known for a single flavor of sterile neutrinos and so can be used both to confirm that the anomaly is truly caused by sterile neutrinos and also to break the degeneracy between the sterile neutrino mixing angle and the sterile neutrino mass splitting.  In the case of C-ADS, obtaining multiple detectors is quite simple.  C-ADS will be located in China's Guangdong province.  In the same province, Daya Bay's eight 20 ton liquid scintillator detectors will no longer be needed.  We therefore suggest using as many of these eight Daya Bay detectors as possible, although in our simulations we have conservatively estimated that only two will be available.  

Not only would the presence of these neutrino detectors be useful for neutrino physics, but the ADS collaboration is also interested in nearby neutrino detectors as, with their reasonably high neutrino event rate, they provide real-time monitoring of the ADS target station and so yield additional information about the reactor as well as a safety measure against, for example, an overheating of the core.

The beam-off backgrounds in this experiment are quite well understood, as they are identical to those in current reactor experiments.  They are essentially all in some way created by cosmogenic muons and so can be considerably reduced with vetoes near these muon events.  We have included the background from decays of spallation ${}^9$Li, which is so long lived that it cannot be effectively removed by muon vetoes at this depth.  We have assumed a depth of 100 meters, which has been considered for the target station.  This gives us 3 background events/detector/day, consistent with that observed by the Daya Bay experiment with a similar overburden.  Although we have run a series of GEANT 4 simulations of the target, described in Ref.~\cite{fengyi}, we have not considered beam-on backgrounds in this analysis, a third Daya Bay detector would be useful to confirm that this approximation is reasonable.   However, with sufficient shielding even the low baseline of 5 m is many times the neutron interaction length and so we expect few neutrons from the target station to arrive at the detector, especially if the detector is placed far from the forward beam axis.

We consider a run with $5\times 10^{23}$ ${}^8$Li decays, each yielding an $\overline{\nu}_e$.  In the absence of sterile neutrinos this yields $2\times 10^4$ IBD events in a 20 ton target volume with 12\% free protons at a distance of 30 meters.  We assume a 5\% uncertainty on this event rate and a 1\% error per 50 keV bin on the background rate.  As we have multiple detectors, the final precision depends only weakly upon this uncertainty unless $\Delta M^2<0.3\ eV^2$, which is well outside the region preferred by global fits of sterile neutrino data.    

Note that, since the ${}^8$Li spectrum is independent of the beam and target station parameters, for any deuterium beam energy and current and any target configuration, one may obtain $5\times 10^{23}$ such decays for an experiment of some duration.   Therefore, given any beam and target parameters, one need only estimate this runtime.   Neutron production rates from 62 MeV and 200 MeV deuteron beams on ${}^9$Be have been measured in Refs.~\cite{deut62,deut200} where  they were found to yield roughly $0.12$ and $0.6$ neutrons/deuteron respectively.   As has been stressed in Ref.~\cite{isodar}, a deuterium beam is advantageous in that it directly brings neutrons to the target, which are liberated via proton stripping and deuterium breakup.  These fast neutrons have energies peaked around half of the deuterium beam energy and so they must travel a long distance before slowing enough to have a high interaction cross section with ${}^7$Li.  This places strong constraints on the target geometry.  For this purpose we have considered a target with multiple ${}^7$Li sleeves surrounded by a graphite reflector.  Our GEANT 4 simulations indicate that nonetheless only about 6\% of the neutrons are captured by ${}^7$Li in our particular target geometry.  This means that we require $8\times 10^{24}$ neutrons and so $1.4\times 10^{25}$ deuterons on target.  With a 15 mA beam this corresponds to 5 years of running.  However an optimization in the target geometry as in Ref.~\cite{isodarussi2015}, so as to increase the neutron capture rate, could easily reduce this runtime.


We approximate the fractional energy resolution of the Daya Bay detectors, 5 years from now, to be
\beq
\sigma=\sqrt{\left(\frac{7\%}{\sqrt{E}}\right)^2+\left(1\%\right)^2}.
\eeq

\begin{figure} 
\begin{center}
\includegraphics[width=2.5in,height=1.0in]{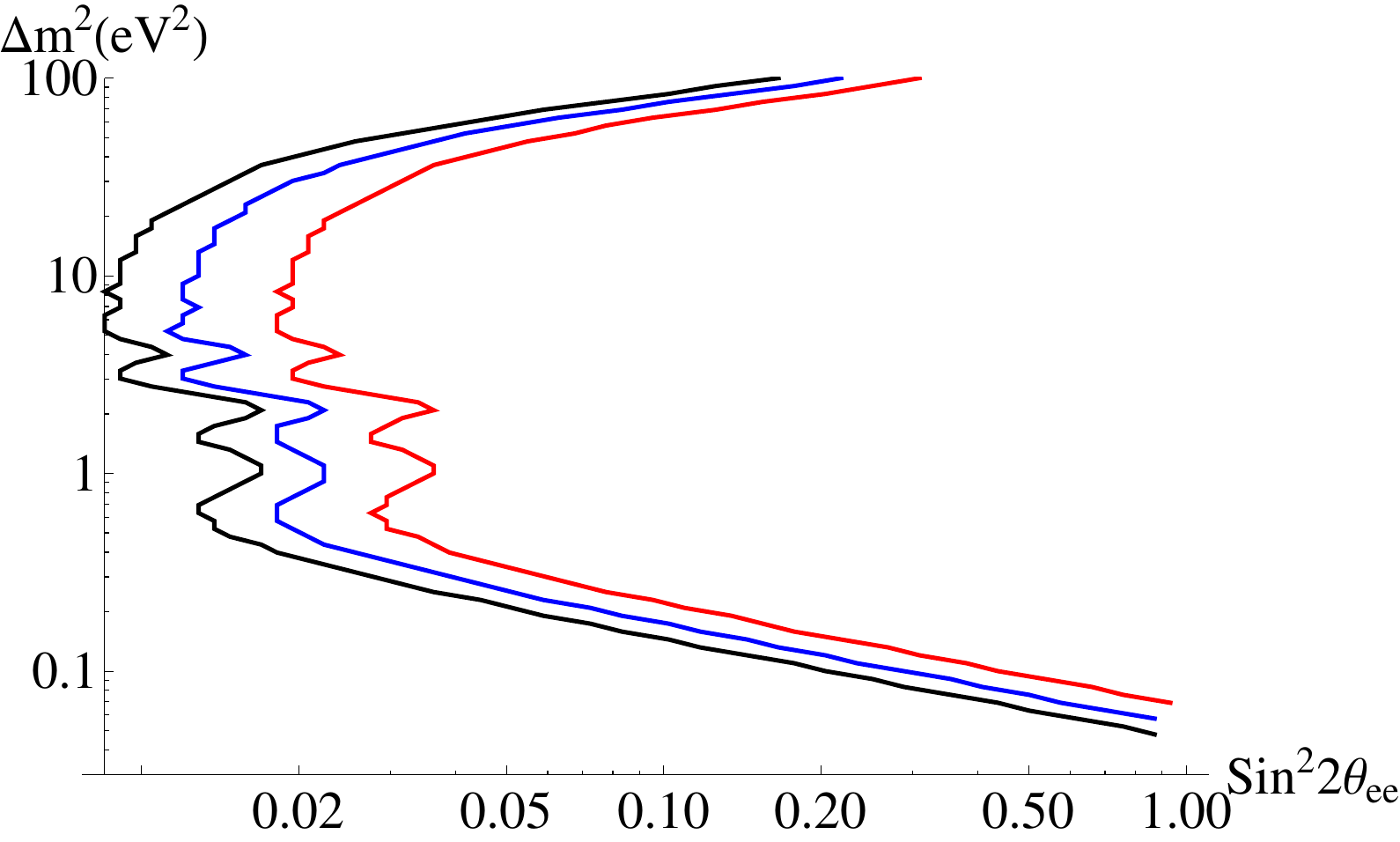}
\caption{The region of $\theta_{ee}-\Delta M^2$ parameter space which may be excluded with 2$\sigma$ (black), 3$\sigma$ (blue) and 5$\sigma$ (red) of confidence by an IsoDAR run with a near detector at 5 m and a far detector at 15 m.}
\label{isofig}
\end{center}
\end{figure}

We have parametrized the $\overline{\nu}_e$ disappearance using the parameters $\theta_{ee}$ and $\Delta M^2$ defined by
\beq
P_{ee}=1-\spp22ee\sin^2\left(1.27\frac{\Delta M^2L}{E}\right)
\eeq
where $L$ is the baseline in meters and $E$ is the neutrino energy in MeV.  We have determined the parameter values which may be excluded at 2$\sigma$, 3$\sigma$ and 5$\sigma$ for all values of pairs of baselines between 5 and 30 meters in a 5 year run, using the crude approximation that the detectors are pointlike.  

Fixing the near and far detector baselines to be 5 m and 15 m, our exclusions are shown in Fig.~\ref{isofig}.  Already in this 2-detector setup, in the 0.5 eV${}^2$-100 eV${}^2$ range this exclusion regime is competitive with the most sensitive proposed sterile neutrino experiments.  This proposal is complimentary to reactor-based experiments that will measure $\theta_{ee}$ because the ${}^8$Li decay spectrum extends about 50\% higher than the relevant part of the reactor spectrum, and so one has access to the same values of $L/E$ with values of $L$ and $E$ which are nonetheless distinct from those so far probed, providing an additional degree of robustness.  With three to eight of the Daya Bay detectors the reach and robustness would be improved considerably.

In the rest of this letter we will consider C-ADS Phase II with a 10 mA beam of 800 MeV protons.   These create neutrinos via $\pi^+$ and $\mu^+$ DAR 
\beq
\pi^+\rightarrow \mu^+ + \nu_\mu\hsp
\mu^+\rightarrow e^+ + \nu_\e + \overline{\nu}_\mu.
\eeq
The $\overline{\nu}_\mu$ spectrum of $\mu$DAR neutrinos is well known.  To determine the normalization of the neutrino signal we again suggest using several near detectors, such as Daya Bay detectors, now at baselines ranging from 30 m to 100 m from the target.  The results that we report below use two Daya Bay detectors, one at 50 m and one at 100 m.  The near detectors can determine the normalization using elastic electron neutrino scattering.  

In addition, the near detectors will again play the dual role of a sterile neutrino search and a real time monitor of the target.  However, the sterile neutrino search will now be most sensitive in the $\overline{\nu}_{\mu}\rightarrow\overline{\nu}_e$ appearance channel.  Thus it is sensitive not to the sterile neutrino mixing angle $\theta_{ee}$ but to $\theta_{\mu e}$, defined by
\beq
P_{\mu e}=\spp22{\mu}e\sin^2\left(1.27\frac{\Delta M^2L}{E}\right)
\eeq
and so it will test of the LSND anomaly \cite{lsnd}.  Like LSND, as an appearance experiment its reach will extend to much lower values of the relevant mixing angle than a disappearance experiment.  In fact, as the accelerator current is an order of magnitude greater than LSND and as there are multiple detectors, it will provide a much more robust test of this anomaly than LSND itself, ranging down to lower mass splittings and smaller mixing angles.

\begin{figure} 
\begin{center}
\includegraphics[width=2.5in,height=1.0in]{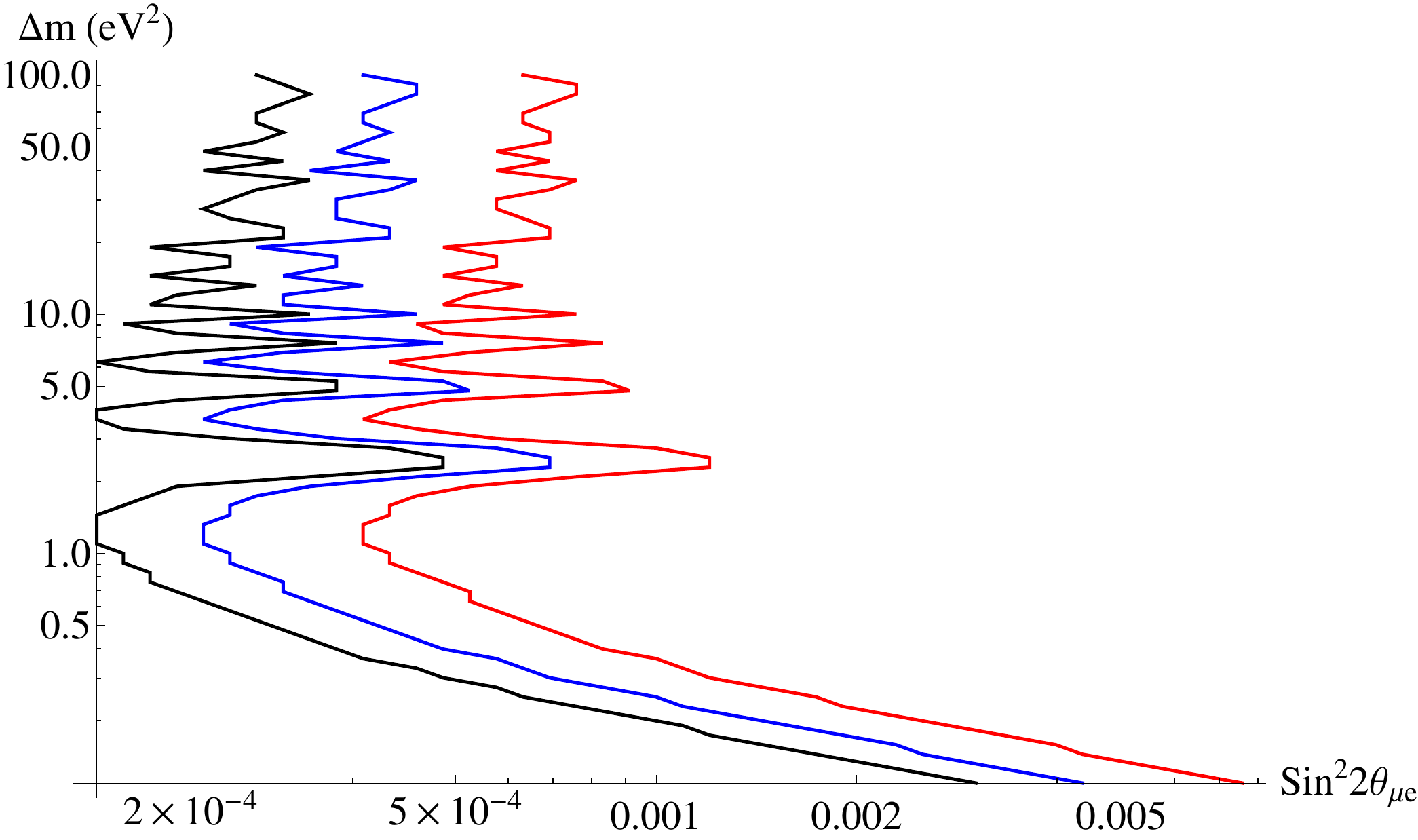}
\caption{The region of $\theta_{\mu e}-\Delta M^2$ parameter space which may be excluded with 2$\sigma$ (black), 3$\sigma$ (blue) and 5$\sigma$ (red) of confidence by a 5 year $\mu$DAR run with a near detector at 50 m and a far detector at 100 m.}
\label{sterilemudarfig}
\end{center}
\end{figure}

The $2\sigma$, $3\sigma$ and 5$\sigma$ bounds which may be achieved on $\theta_{\mu e}$ and $\Delta M^2$ in a 5 year run are presented in Fig.~\ref{sterilemudarfig}.  Note that the best fit LSND anomaly region is well covered, and so this will provide a robust test of the LSND anomaly.  Elastic $e-\nu$ scattering in the near detectors can be used to determine the $\overline{\nu}_\mu$ flux with an uncertainty of about 5\%, which is essentially irrelevant for such an appearance channel sterile neutrino search.  The beam-off backgrounds are dominated by atmospheric neutrinos which will be relevant at the far detectors but at the near detectors yield backgrounds which are again negligible.  The beam-on backgrounds are more difficult to anticipate, but little is expected in this energy range.  Following the estimates given by LSND \cite{lsnd} we fix the beam-on background by asserting that for each $\mu^+$ DAR there are $5\times 10^{-4}$ $\mu^-$ DARs and that the normalization of this background is known with a precision of 25\%.  We find that this beam-on background has little effect on our results.  If there does happen to be an unexpected beam-on background, its contribution to the appearance rate at the near detectors would scale as $1/r^2$, in contrast with the sterile neutrino appearance signal, and so the two could in principle be disentangled.  Of course in such a case it would be reasonable to bring more Daya Bay near detectors at different baselines to distinguish between the two signals as robustly as possible.

So far the proposed physics program has been quite cheap, as the $\overline{\nu}$ have been supplied for free by ADS, albeit with a modified target station in the first setup, and the detectors have been recycled from Daya Bay.  However perhaps the most exciting potential application of ADS reactor neutrinos is the measurement of the leptonic CP-violating phase $\delta$.   For this purpose, we will consider two 20 kton liquid scintillator detectors at baselines of 2.5-30 km to detect $\overline{\nu}_e$ created by $\overline{\nu}_\mu$ oscillations.  Combining this appearance result with $\nu_\mu\rightarrow\nu_e$ from NO$\nu$A and T2K one obtains the leptonic CP-violating phase $\delta$.  Unlike the 20 kton detectors needed for mass hierarchy experiments, the measurement of $\delta$ does not impose stringent restrictions upon the detector's energy resolution or the knowledge of its nonlinear energy response.  Thus, unlike JUNO \cite{juno}, the detector can be cylindrical, the PMT coverage can be less than 50\% and the scintillator purity does not need to exceed that in present day experiments.  Also, as 80\% of the events are at energies above 30 MeV and so well above the spallation isotope backgrounds, the detectors do not need to be nearly as deep as JUNO.

Such a setup has numerous advantages for measuring $\delta$ \cite{noidarts,kaoru}.  Not only is the $\overline{\nu}_e$ spectrum well known, but since they are detected via IBD the cross section is also well known and the capture of the resulting neutron provides a double coincidence which can be used for background rejection.  As a $\overline{\nu}$ oscillation experiment it maximizes synergy with beam experiments such as T2K and NO$\nu$A, which are most efficient in neutrino mode.   The $\overline{\nu}$ energies are higher than geoneutrino, reactor neutrino or even spallation neutrino energies, and yet low enough so that atmospheric neutrino backgrounds are small.  Unlike beam experiments the $\overline{\nu}$ energy is also below the muon mass, and so there is no background from charged current $\nu_\mu$ or $\overline{\nu}_\mu$ interactions.   The spectrum is broad enough that such experiments can cleanly distinguish $\delta$ from $\pi-\delta$.  Finally the main background that appears in water Cherenkov $\mu$DAR experiments \cite{daed,kaoru}, the invisible muons, are not present in a scintillator detector as even low energy muons, via ionization, emit scintillation light.

\begin{figure} 
\begin{center}
\includegraphics[width=2.5in,height=1.0in]{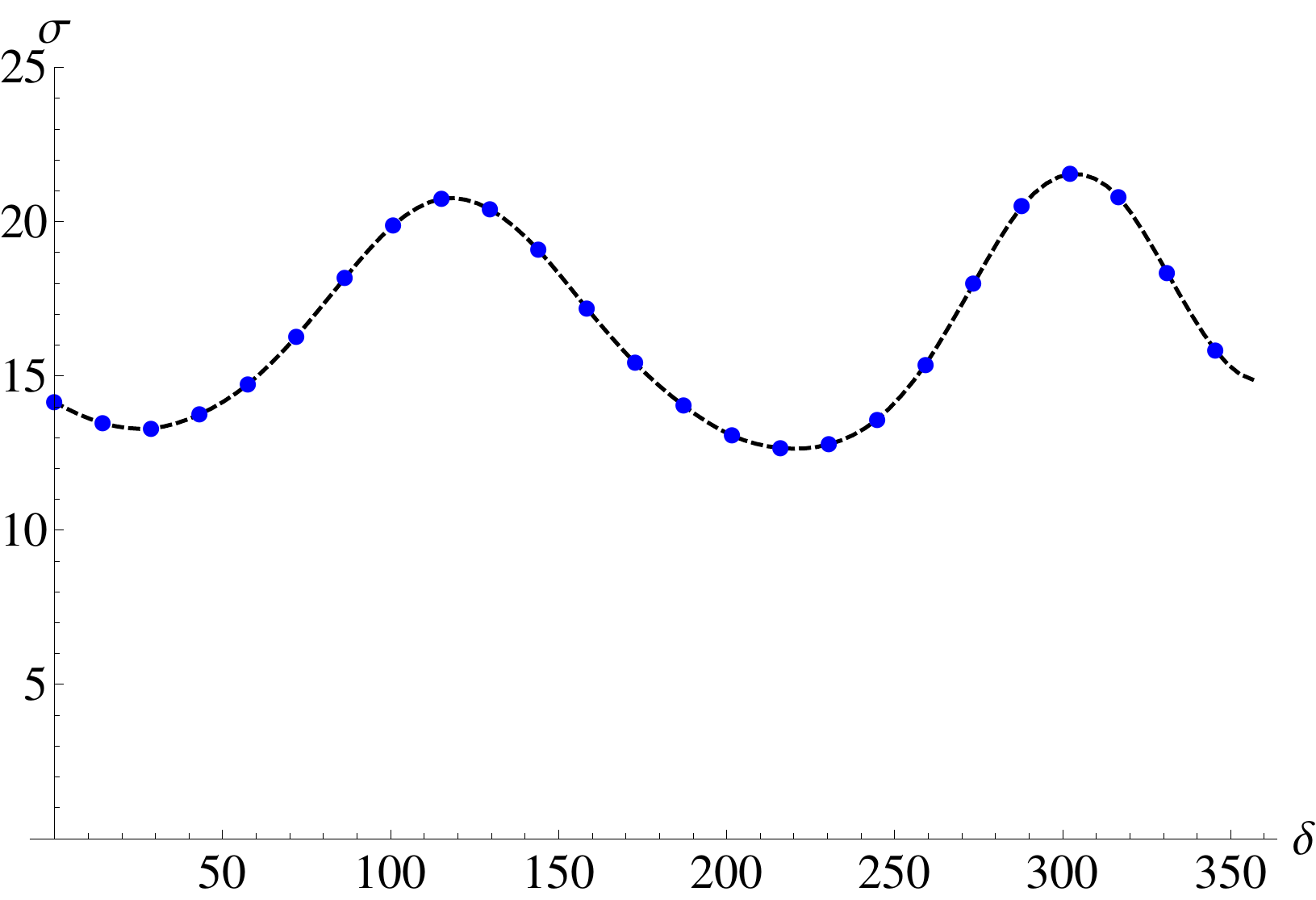}
\caption{The 1$\sigma$ precision with which $\delta$ can be measured, for each true value of $\delta$, by a 12 year $\mu$DAR run with Daya Bay near detectors and 20 kton liquid scintillator detectors at 5 km and 25 km.  12 years of appearance data at NO$\nu$A is also used. The dots are our numerical results, whereas the curve is an interpolation.}
\label{duerivelfig}
\end{center}
\end{figure}

\begin{figure} 
\begin{center}
\includegraphics[width=2.5in,height=1.0in]{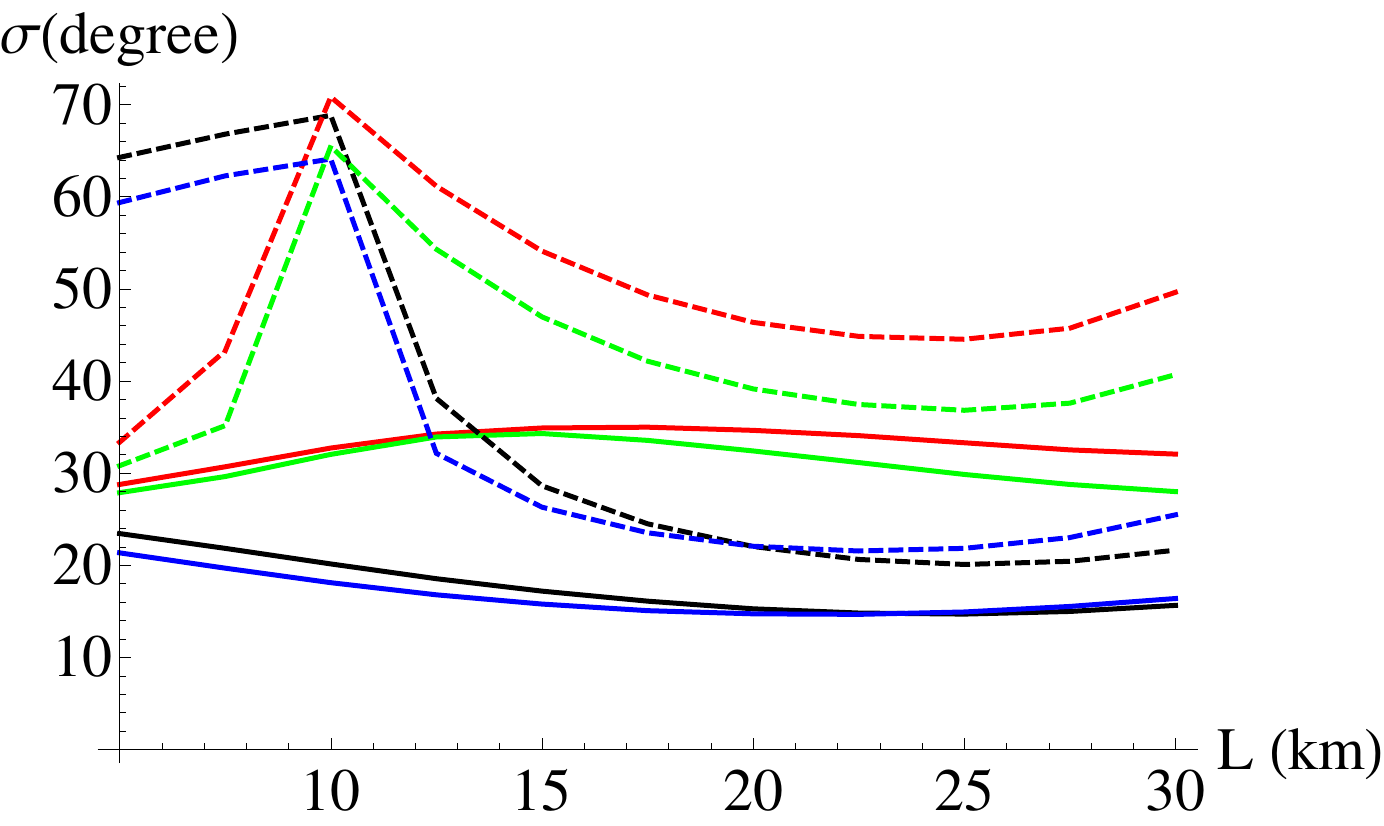}
\caption{The $\delta$-averaged 1$\sigma$ precision with which $\delta$ can be measured by $\mu$DAR with a {\it{single}} 20 kton detector at the given baseline if the true value of $\delta$ is $0^\circ$ (black), $90^\circ$ (red), $180^\circ$ (blue) or $270^\circ$ (green).  Solid curves include NO$\nu$A.  }
\label{unorivelfig}
\end{center}
\end{figure}

We have considered a 12 year $\mu$DAR run together with 12 years of NO$\nu$A appearance data, half in $\nu$ mode and half in $\overline{\nu}$ mode, defined as in Ref.~\cite{nova}.  The main background comes from atmospheric $\nu_e$ and $\overline{\nu}_e$ interacting via IBD and also charged current quasielastic interactions.  Roughly adapting the results of the simulations in Ref.~\cite{kaoru}, we determine the shape of this background and estimate the total number of events to be about 50 in a 12 year run in southern China.  In other locations the background rate will be higher, as the strong horizontal component of the magnetic field in southern China leads to a smaller low energy neutrino flux.  With two 20 kton detectors, one at 5 km and another at 25 km, the precision with which $\delta$ can be measured is given in Fig.~\ref{duerivelfig}.  The precision, about $15^\circ$, is competitive with other future proposals such as DUNE and Hyper-Kamiokande but at a lower cost.   Even with a single detector, as can be seen in Fig.~\ref{unorivelfig}, a precision of about $25^\circ$ can be expected.  Needless to say, a single detector can be built first and the second detector can be built later if and when the necessary funds present themselves.

\section* {Acknowledgement}
\noindent
JE and EC are supported by NSFC grant 11375201.  EC  is also supported by the Chinese Academy of Sciences President's International Fellowship Initiative grant 2015PM063. 


\end{document}